\documentclass[conference]{IEEEtran}
\IEEEoverridecommandlockouts
\usepackage{cite}
\usepackage{amsmath,amssymb,amsfonts}
\usepackage{algorithm}
\usepackage{algorithmicx}
\usepackage{algpseudocode}

\usepackage{graphicx}
\usepackage{textcomp}
\usepackage{xcolor} 
\def\BibTeX{{\rm B\kern-.05em{\sc i\kern-.025em b}\kern-.08em
    T\kern-.1667em\lower.7ex\hbox{E}\kern-.125emX}}
 
\usepackage{enumitem}
\usepackage[utf8]{inputenc}
\usepackage{amsfonts}
\usepackage{mathtools}
\mathtoolsset{showonlyrefs}
\usepackage{xcolor,xspace}
\usepackage{setspace}
\usepackage{theorem}
\usepackage{pgfplots}
\usepackage{mathrsfs}
\usetikzlibrary{plotmarks}

\usepackage{amssymb}

\definecolor{TUMdgray}{RGB}{088,088,090}
\definecolor{TUMmgray}{RGB}{156,157,159}
\definecolor{TUMlgray}{RGB}{215,217,218}
\definecolor{TUMyellow}{RGB}{255,180,000}
\definecolor{TUMorange}{RGB}{255,128,000}
\definecolor{TUMblue}{RGB}{000,101,189}
\definecolor{TUMgreen}{RGB}{0,124,48}
\definecolor{TUMred}{RGB}{196,7,27}
\definecolor{TUMgold}{RGB}{255,200,0}

\pgfplotsset{
	compat=1.14,
    every axis/.append style={
        scale only axis,
	width=0.8\textwidth,
	height=0.8\textwidth,
	label style={inner sep=0, font=\normalsize}, 
	tick label style={font=\scriptsize},
	legend style={font=\scriptsize},
	mark size=3,
	major grid style={dashed},
	line width=0.8pt,
	axis line style = thin}
}

\newtheorem{theorem}{Theorem}
\newtheorem{definition}{Definition}
\newtheorem{lemma}{Lemma}

\newtheorem{proposition}{Proposition}

\newcommand{\Fqm}{\ensuremath{\mathbb F_{q^m}}}

\newcommand{\Fq}{\ensuremath{\mathbb F_{q}}}

\newcommand{\F}{\ensuremath{\mathbb F}}

\newcommand{\set}[1]{\ensuremath{#1}}

\newcommand{\intervallincl}[2]{\ensuremath{[#1,#2]}}

\newcommand{\spannedBy}[1]{\ensuremath{\langle #1\rangle_q}}

\DeclareMathOperator{\defi}{def}
\newcommand{\defeq}{\overset{\defi}{=}}

\DeclareMathOperator{\rk}{rk}

\DeclareMathOperator{\supp}{supp}

\renewcommand{\vec}[1]{\ensuremath{\mathbf{#1}}}
\newcommand{\mat}[1]{\ensuremath{\mathbf{#1}}}

\renewcommand{\a}{\mathbf a}
\renewcommand{\b}{\mathbf b}

\newcommand{\e}{\mathbf e}

\newcommand{\s}{\mathbf s}

\newcommand{\A}{\mathbf A}
\newcommand{\B}{\mathbf B}

\newcommand{\D}{\mathbf D}

\renewcommand{\H}{\mathbf H}

\newcommand{\0}{\mathbf 0}

\newcommand{\mycode}[1]{\ensuremath{\mathcal{#1}}}

\newcommand{\myspace}[1]{\mathcal{#1}}

\newcommand{\Aspace}{\myspace{A}}
\newcommand{\Bspace}{\myspace{B}}

\newcommand{\Cint}{\mathcal{C}_{\text{int}}}
\newcommand{\Clong}{\mathcal{C}_{\text{lon}}}

\newcommand{\dR}{d_{\text{R}}}

\newcommand{\Hspace}{\myspace{F}}
\newcommand{\Hdim}{\lambda}
\newcommand{\Hb}{\varphi}
\newcommand{\Hset}{\Phi}

\newcommand{\Espace}{\myspace{E}}
\newcommand{\Edim}{t}
\newcommand{\Eb}{\gamma}
\newcommand{\Eset}{\Gamma}

\newcommand{\Pspace}{\mathcal {P}}

\newcommand{\Sspace}{\myspace{S}}

\newcommand{\SspaceInt}{\myspace{S}^{\prime}}

\renewcommand{\u}{u}
\newcommand{\w}{w}

\newcommand{\Gint}{\vec{G}_{\text{int}}}
\newcommand{\Gcomp}{\vec{G}_{\text{c}}}

\newcommand{\Solve}{{\tt Solve}}

\begin{document}
\title{Efficient Decoding of \\ Interleaved Low-Rank Parity-Check Codes\\  
  \thanks{This project has received funding from the European Research Council (ERC) under the European Union’s Horizon 2020 research and innovation programme (grant agreement No~801434)
  }
}

\author{\IEEEauthorblockN{Julian Renner}
\IEEEauthorblockA{\textit{Institute for Communications Engineering} \\
\textit{Technical University of Munich (TUM)}\\
Munich, Germany \\
julian.renner@tum.de}
\and
\IEEEauthorblockN{Thomas Jerkovits, Hannes Bartz}
\IEEEauthorblockA{\textit{Institute of Communications and Navigation} \\
\textit{German Aerospace Center (DLR)}\\
Oberpfaffenhofen-Wessling, Germany \\
\{thomas.jerkovits, hannes.bartz\}@dlr.de}
}

\maketitle

\begin{abstract}
An efficient decoding algorithm for horizontally \textit{u}-interleaved LRPC codes is proposed and analyzed. 
Upper bounds on the decoding failure rate and the computational complexity of the algorithm are derived. 
It is shown that interleaving reduces the decoding failure rate exponentially in the interleaving order \textit{u} whereas the computational complexity grows linearly. 
\end{abstract}


\section{Introduction}

Rank-metric codes have been introduced independently in~\cite{Gabidulin_TheoryOfCodes_1985,Delsarte_1978,Roth_RankCodes_1991} and are used e.g. for network coding~\cite{silva2008rank,sidorenko2010decoding} and for constructing space-time codes (see e.g.~\cite{gabidulin2000space}).
The generic decoding problem in the rank metric is much harder than in the Hamming metric which makes rank-metric codes good candidates to design quantum-resistant code-based cryptosystems~\cite{gaborit2013LRPC,Gabidulin2003Reducible,loidreau2016evolution,faure2006new,wachter2018repairing}. 

The most famous class of rank-metric codes are Gabidulin codes which achieve the Singleton-like bound for rank-metric codes and thus are called maximum rank distance (MRD) codes~\cite{Gabidulin_TheoryOfCodes_1985}.
However, most variants of Gabidulin code-based cryptosystems suffer from structural attacks due to the inherent code structure.

Low-Rank Parity-Check (LRPC) codes are another class of rank-metric codes and have been proposed by Gaborit et al.~\cite{gaborit2013LRPC}. Similar to Low-Density Parity-Check codes, LRPC codes are generated in a randomized way and their decoding is probabilistic with some residuent \emph{decoding failure rate} (DFR).
Although the error-correction performance of LRPC codes is worse compared to Gabidulin codes, LRPC codes are good candidates for designing code-based cryptosystems since they are highly unstructured. 

Ideal LRPC codes have been proposed to reduce the key size of LRPC code-based cryptosystems by adjusting the trade-off between structure in the code and security constraints~\cite{gaborit2019LRPC}.

LRPC codes are used e.g. in the ROLLO cryptosystem~\cite{rollo2019} which is a current candidate in the second round of the NIST standardization process for quantum-resistant cryptosystems. 
 
In this paper, we consider horizontally $\u$-interleaved LRPC codes, which are obtained by the $\u$-fold Cartesian product of an LRPC component code.
The resulting code has length $\u n$ and dimension $\u k$, where $n$ and $k$ are the length and the dimension of the component code, respectively.
This LRPC code construction is motivated by the difficult generic decoding problem stated in~\cite[Definition 7]{gaborit2017IBE}.

An efficient decoding algorithm for horizontally $\u$-interleaved LRPC codes is proposed and analyzed. 
The DFR of the proposed algorithm decreases exponentially in the interleaving order $\u$ whereas the computational complexity increases \emph{linearly} in $\u$. 
We observe that although an LRPC code of length $\u n$ has the same error correction capability as a $\u$-interleaved LRPC code of the same rate, interleaved LRPC codes benefit from 1) a $\u$-times lower decoding complexity and 2) a more compact representation of the code which allows to reduce the key size of LRPC code-based cryptosystems.


\section{Preliminaries}

Let $\Fq$ be a finite field of order $q$ and denote by $\Fqm$ the extension field of $\Fq$ of degree $m$.
The set of all vectors of length $n$ with elements from $\Fq$ is denoted by $\Fq^n$.
By fixing a basis of $\Fqm$ over $\Fq$ each element from $\Fqm$ can be uniquely represented by a vector from $\Fq^m$, i.e. there is a bijective mapping from $\Fqm$ to $\Fq^m$.
By $\intervallincl{1}{n}$ we denote the set of integers $\{1,2,\dots,n\}$.
Matrices and vectors are denoted by bold upper-case and lower-case letters such as $\mat{A}$ and $\vec{a}$, respectively.
The elements of vectors and matrices are indexed beginning from one, like e.g.
\begin{equation}
	\mat{A}=
	\begin{pmatrix}
	 a_{1,1} & a_{1,2} & \dots & a_{1,n}
	 \\
	 \vdots & \vdots & \ddots & \vdots
	 \\
	 a_{m,1} & a_{m,2} & \dots & a_{m,n}
	\end{pmatrix}.
\end{equation}
The rank norm $\rk_q(\a)$ of a vector $\a\in\Fqm^n$ is the rank of the matrix representation $\A \in \Fq^{m \times n}$ over $\mathbb{F}_{q}$, i.e.,
\begin{equation*}
\rk_q(\a) := \rk_q(\A).
\end{equation*}

Given a set $\set{A}=\left\{a_1,a_2,\dots,a_{n}\right\}\subseteq\Fqm$ we denote by \spannedBy{\set{A}} the $\Fq$-linear subspace spanned by the elements in $\set{A}$.
The support of a vector $\a\in\Fqm^n$ is defined as $\supp(\a) \defeq \spannedBy{a_1,\hdots,a_n}$.

\subsection{Rank-Metric Codes}

The rank distance between two vectors $\a$ and $\b$ is defined as
\begin{equation*}
\dR(\a,\b):= \rk_q(\a-\b) = \rk_q(\A-\B)
\end{equation*}
where $\A$ and $\B$ are the matrix representations of $\a$ and $\b$, respectively.

A linear $[n,k,d]$ rank-metric code $\mycode{C}$ of length $n$, dimension $k$ and minimum rank distance $d$ over $\Fqm$ is a $k$-dimensional subspace of $\Fqm^n$ where
\begin{equation*}
d := \min_{\substack{\a,\b \in \mycode{C} \\ \a \neq \b }} \lbrace \rk_q(\a -\b) \rbrace  =  \min_{\a \in \mycode{C} \setminus \{0\} }\lbrace \rk_q(\a) \rbrace. 
\end{equation*}

The codewords $\vec{c}\in\mycode{C}$ of a rank-metric code $\mycode{C}$ are transmitted over a channel
\begin{equation}
	\vec{y}=\vec{c}+\vec{e},
\end{equation}
where the rank of the error vector $\vec{e}$ is equal to $\Edim$.
Let $\Espace=\supp(\vec{e})$ be the support of the error vector $\vec{e}$ and let $\Eset=\{\Eb_1,\Eb_2,\dots,\Eb_{\Edim}\}\subset{\Fqm}$ denote a basis for $\Espace$.
Then each entry $e_j$ of the error vector $\vec{e}$ can be written as an $\Fq$-linear combination 
\begin{equation}\label{eq:eExp}
 e_{j}=\sum_{r=1}^{\Edim}e_{j,r}\Eb_{r},
 \qquad j\in\intervallincl{1}{n}
\end{equation}
for some elements $e_{j,r}$ from $\Fq$.


\section{LRPC Codes and their Decoding}
In this section, we give a brief overview on LRPC codes and the efficient decoding algorithm from~\cite{gaborit2013LRPC,gaborit2019LRPC}.

\subsection{Low-Rank Parity-Check Codes}
\begin{definition}[Low-Rank Parity-Check Code]\label{def:LRPC}
 An LRPC code $\mycode{C}[\Hdim;n,k]$ of length $n$, dimension $k$ and rank $\Hdim$ over $\Fqm$ is defined as a code with a parity-check matrix $\mat{H}\in\Fqm^{(n-k)\times n}$ where the vector space $$\Hspace=\spannedBy{\{h_{i,j}:i\in [1,n-k],j\in[1,n]\}}$$ has dimension at most $\Hdim$.
\end{definition}
Let the set $\Hset = \{\Hb_1,\Hb_2,\dots,\Hb_{\Hdim}\}$ be a basis for $\Hspace$.
Then we can write each element of the matrix $\mat{H}$ as an $\Fq$-linear combination of the basis elements in $\Hset$, i.e. we have
\begin{equation}\label{eq:hExp}
	h_{i,j}=\sum_{\ell=1}^{\Hdim}h_{i,j,\ell}\Hb_{\ell},
	\qquad 
	\forall i\in\intervallincl{1}{n-k},j\in\intervallincl{1}{n}
\end{equation}
for some $h_{i,j,\ell}\in\Fq$.

\subsection{Decoding of LRPC Codes}

The basic decoding algorithm for LRPC codes in~\cite{gaborit2013LRPC,gaborit2019LRPC} uses properties of product spaces, which are defined as follows.

\begin{definition}[Product Space]
  Let $\Aspace$ and $\Bspace$ be two $\Fq$-subspaces of $\Fqm$. 
  Then the product space $\Aspace \Bspace$ of $\Aspace$ and $\Bspace$ is defined as 
  \begin{equation}
  	\Aspace \Bspace\defeq \spannedBy{\{ab:a\in\Aspace,b \in \Bspace \}}.
  \end{equation}
\end{definition}

Note, that the dimension of the product space $\Aspace \Bspace$ is bounded from above by $\dim(\Aspace \Bspace)\leq\dim(\Aspace)\dim(\Bspace)$.

The decoding algorithm consists of three parts and can be summarized as follows.
\begin{enumerate}[leftmargin=*]
\item \textbf{Compute the syndrome space:} \\Compute the syndrome vector $\vec{s} \defeq \vec{y} \H^{\top}$ and the determine a basis of the syndrome space $\Sspace \defeq \supp(\s)$.
\item \textbf{Recover the support of the error:} \\ Compute a basis for all $\Sspace_{\ell} \defeq \Hb_{\ell}^{-1} \Sspace$, $\ell \in [1,\Hdim]$, estimate the support of the error by $\Espace = \Sspace_{1} \cap \hdots \cap \Sspace_{\Hdim}$ and determine a basis $\Eset$ of $\Espace$.
\item \textbf{Recover the error vector:} \\
In order to recover the error vector $\e$ from the support $\myspace{E}$ we need to compute the elements $e_{j,r}$ in~\eqref{eq:eExp}.
This can be done efficiently by solving 
\begin{equation}\label{eq:solveError}
s_{i,\ell,r} =\sum_{j=1}^{n}h_{i,j,\ell} e_{j,r},\quad
 \ell\in[1,\Hdim], r\in[1,\Hdim], i\in[1,n-k]
\end{equation}
where $s_{i,\ell,r}$ is the expansion w.r.t. to the product space basis such that $s_{i}=\sum_{r=1}^{\Edim}\sum_{\ell=1}^{\Hdim}s_{i,\ell,r}\Hb_\ell\Eb_r$.
Note that~\eqref{eq:solveError} is a linear system of $(n-k)\Hdim\Edim$ equations in $n\Edim$ unknowns.

\begin{proposition}
  The linear system \eqref{eq:solveError} has a unique solution if the matrix $\H_{\text{ext}}\in \Fq^{(n-k)\Hdim \times n}$ defined as
\begin{equation*}
 \H_{\text{ext}} \defeq
 \begin{pmatrix}
   h_{111} & h_{121} & \hdots & h_{1n1} \\
   h_{112} & h_{122} & \hdots & h_{1n2} \\
   \vdots & \vdots & \ddots & \vdots \\
  h_{11\Hdim } & h_{12\Hdim} & \hdots & h_{1n\Hdim} \\
  h_{211} & h_{221} & \hdots & h_{2n1} \\
  h_{212} & h_{222} & \hdots & h_{2n2} \\
  \vdots & \vdots & \ddots & \vdots \\
  h_{(n-k)1\Hdim} & h_{(n-k)2\Hdim} & \hdots & h_{(n-k)n\Hdim} \\
\end{pmatrix}
\end{equation*}
has full rank.
\end{proposition}
\end{enumerate}

\subsection{Upper Bounds on the Decoding Failure Rate}

For unfortunate choices of the parity-check matrix $\H$, we may have $\rk(\H_{\text{ext}}) < n$ which makes the decoder fail for any error pattern.
Since $\H_{\text{ext}}\in\Fq^{(n-k)\Hdim \times n}$ we can have $\rk(\H_{\text{ext}}) = n$ only if $\Hdim \geq \frac{n}{n-k}$.  
If $\Hdim \geq \frac{n}{n-k}$ and we choose $\mycode{C}$ uniformly at random among all $[\Hdim;n,k]$ LRPC codes, the probability that $\rk(\H_{\text{ext}}) < n$ is equal to the probability that an $(n-k)\Hdim \times n$ matrix over $\Fq$ is not full rank, i.e.,
\begin{equation*}
\Pr[\rk(\H_{\text{ext}}) < n] = 1- \frac{1}{q^{(n-k)\Hdim n}} \prod_{j=1}^{n} \Big(q^{(n-k)\Hdim}-q^{j-1}\Big),
\end{equation*}
see~\cite{Migler2004RankMatrix}. 
This event depends on the particular choice of $\H$ and can be avoided by drawing $\H$ randomly until $\H_{\text{ext}}$ has full rank.

There are three cases in which the LRPC decoding algorithm fails:
\begin{enumerate}
\item $\dim(\Hspace \Espace) < \Hdim\Edim$,
\item $\Espace \neq \Sspace_{1} \cap \Sspace_{2} \cap \hdots \cap \Sspace_{\Hdim}$,
\item $\Sspace \subsetneq \Hspace \Espace$.
\end{enumerate}

Upper bounds on the probability of the failure events are derived in~\cite{gaborit2019LRPC} and shown in Table~\ref{tab:failureLRPC}. For most practical parameters the event $\Sspace \subsetneq \Hspace \Espace$ dominates the DFR. 

\begin{table}
\caption{Upper Bounds on the Decoding Failure Probabilities~\cite{gaborit2019LRPC}}
\renewcommand{\arraystretch}{1.3} 
\begin{center}
\begin{tabular}{r|l}
Event & Failure Probability \\
  \hline
  $\dim(\Hspace \Espace) < \Hdim\Edim$ & $\leq \Edim q^{\Hdim\Edim -m}$ \\
  $ \Sspace_{1} \cap \Sspace_{2} \cap \hdots \cap \Sspace_{\Hdim} \neq \Espace $& $\leq \Edim q^{0.5\Hdim\Edim(\Hdim+1)-m}$ \\
  $\Sspace \subsetneq \Hspace \Espace$ & $  \leq  q^{\Hdim\Edim - (n-k)}$
\end{tabular}
\end{center}
\label{tab:failureLRPC}
\end{table}

\subsection{Complexity of the Decoding Algorithm}

An estimate of the computational complexity of the LRPC decoding algorithm is given in~\cite{gaborit2019LRPC}. 
In the following, we provide a detailed complexity analysis.

\begin{lemma}
  The algorithm presented in~\cite{gaborit2019LRPC} requires
$O(n^2m^2)$
operations in $\Fq$ to decode a $[\Hdim;n,k]$ LRPC code over $\Fqm$.
  \end{lemma}
\begin{IEEEproof}
The first step consists of computing the syndrome vector and determining a basis of the product space, which can be achieved by transforming an $m \times (n-k)$ matrix over $\Fq$ in reduced row echelon form. Computing the syndrome has complexity of $O((n-k)n)$ operations in $\Fqm$ or $O((n-k)n m^2)\subset O(n^2m^2)$ operations in $\Fq$~\cite[Remark 8]{puchinger2019HighOrderInt}, and the transformation in reduced row echelon form requires $O(\min\{m^2(n-k);m(n-k)^2 \})\subset O(\min\{m^2 n;m n^2\})$. Thus for the first step, we require $O(n^2 m^2)$ operations in $\Fq$.

Recovering the support of the error in the second step requires $O(4\Edim^2\Hdim^2m)$ operations in $\Fq$~\cite[Section 4.5]{gaborit2019LRPC}.

The final step can be performed by solving a linear system of $(n-k)\Hdim\Edim$ equations and $n\Edim $ unknowns over $\Fq$. This requires $O((n-k)^2\Hdim^2\Edim^2n\Edim) \subset O(n^3\Edim^3\Hdim^2)$ operations over $\Fq$.
Alternatively, we can precompute a matrix $\D_{H} \in \Fq^{n\Edim\times n \Edim}$, as described in~\cite[Section 4.5]{gaborit2019LRPC}, and perform the final step by a single matrix--vector multiplication which requires $O(n^2 \Edim^2)$ operations over $\Fq$.
\end{IEEEproof}


\section{Interleaved LRPC Codes and their Decoding}
In this section, we introduce (horizontally) interleaved LRPC codes and propose an efficient decoding algorithm.
\subsection{Interleaved Low-Rank Parity-Check Codes}
\begin{definition}[Interleaved Low-Rank Parity-Check Code]\label{def:intLRPC}
  Let $\mycode{C}[\Hdim;n,k]$ be an LRPC code of length $n$, dimension $k$ and rank $\Hdim$ as in Definition~\ref{def:LRPC}. The corresponding (horizontally) $\u$-interleaved LRPC code $\mycode{IC}[\u,\Hdim;n,k]$ is defined as
  \begin{equation}
  \{ \big(\vec{c}^{(1)},\hdots,\vec{c}^{(\u)}\big)\in\Fqm^{\u n}: \vec{c}^{(w)} \in \mycode{C}[\Hdim;n,k], \forall w\in\intervallincl{1}{\u} \}.
    \end{equation}
\end{definition}
A $\u$-interleaved LRPC code $\mycode{IC}[\u,\Hdim;n,k]$ has length $\u n$ and dimensions $\u k$ over $\Fqm$. 
By using ideal LRPC codes as component codes $\mycode{C}[\Hdim;n,k]$ we can construct $\u$-interleaved ideal LRPC codes (see \cite[Defintion 4.2]{gaborit2019LRPC}).
In this paper we focus on ordinary interleaved LRPC codes. 

\subsection{Decoding of Interleaved LRPC Codes}
Suppose a codeword of a horizontally $\u$-interleaved LRPC code $\mycode{IC}[\u,\Hdim;n,k]$ is transmitted and
\begin{align}
  \vec{y} &= \big(  \vec{y}^{(1)},\hdots,  \vec{y}^{(\u)} \big) \\
  &= \big(\vec{c}^{(1)},\hdots,\vec{c}^{(\u)}\big) + \big(\vec{e}^{(1)},\hdots,\vec{e}^{(\u)}\big)
\end{align}
is received. 
In contrast to independent transmissions of the $\u$ component codewords over a rank error channel of rank $t$, the component error vectors $\vec{e}^{(1)},\hdots,\vec{e}^{(\u)}$ share the same support $\Espace$, i.e., we have 
\begin{equation}
\spannedBy{e_{1}^{(1)},\hdots,e_{n}^{(1)} \ | \ \hdots \ | \ e_{1}^{(\u)},\hdots,e_{n}^{(\u)}} = \Espace
\end{equation}
and $\dim(\Espace)=\Edim$.
Using~\eqref{eq:eExp} we can write the entries of the $\w$-th component error as
\begin{equation}
e^{(\w)}_j=\sum_{r=1}^{\Edim} e_{j,r} \Eb_{r}, \qquad \forall j \in [1,n].
\end{equation}
Defining the matrices
\begin{equation}
	\mat{E}^{(\w)}=
	\begin{pmatrix}
	 e_{1,1}^{(\w)} & e_{1,2}^{(\w)} &  \dots & e_{1,\Edim}^{(\w)}
	 \\
	 \vdots & \dots & \ddots & \vdots
	 \\
	 e_{n,1}^{(\w)} & e_{1,2}^{(\w)} & \dots & e_{n,\Edim}^{(\w)}
	\end{pmatrix}
	\in\Fq^{n\times\Edim},\quad \forall \w\in[1,\u]
\end{equation}
we can write the interleaved error vector $\vec{e}$ as
\begin{equation}\label{eq:eExpInt}
\vec{e}=\left(\Eb_1,\Eb_2,\dots,\Eb_\Edim\right)
\begin{pmatrix}
 \mat{E}^{(1)\top} & \mat{E}^{(2)\top} & \dots & \mat{E}^{(\u)\top}
\end{pmatrix}.
\end{equation}

In the following, we present a decoding algorithm for interleaved LRPC codes. Similar to the non-interleaved case, the algorithm consists of three steps:
\smallskip
\begin{enumerate}[leftmargin=*]
\item \textbf{Computation of the syndrome space,}
\item \textbf{Recovery of the support of the error,}
\item \textbf{Recovery of the error vector.} 
\end{enumerate}
\smallskip

We analyze the three steps and derive upper bounds on their failure probabilities.

\subsubsection{Computation of the Syndrome Space}
In the first step, we determine the space that is spanned by the entries of the syndromes $\vec{s}^{(1)},\dots,\vec{s}^{(\u)}$, i.e. $\SspaceInt\defeq\supp{\left((\s^{(1)}, \s^{(2)}, \dots, \s^{(\u)})\right)}$.
\begin{lemma}
  Let $ \Pspace \defeq \Hspace\Espace$ and let $\vec{s}^{(\w)} = \vec{y}^{(\w)}\mat{H}^{\top}$. 
  Then we have
  \begin{equation*}
	\SspaceInt=\supp{\left((\s^{(1)}, \s^{(2)}, \dots, \s^{(\u)})\right)} \subseteq \Pspace.
  \end{equation*}
  Under the assumption that $\dim(\myspace{P})=\lambda t$ we have that
\begin{equation*}
\Pr[ \Sspace \subsetneq \Pspace ] \leq q^{\Hdim\Edim - \u(n-k)}.
\end{equation*}
\label{lemma:comp_syndrome}
\end{lemma}
\begin{IEEEproof}
The $\w$-th syndrome $\vec{s}^{(\w)}\in\Fqm^{n-k}$ is computed as 
\begin{equation}
	\vec{s}^{(\w)}=\vec{y}^{(\w)}\mat{H}^{\top} = \vec{e}^{(\w)}\mat{H}^{\top},
\end{equation}
where each entry can be written as
\begin{equation}\label{eq:IntSyndromeEntry}
	s_{i}^{(\w)}=\sum_{j=1}^{n}h_{i,j}e_{j}^{(\w)}.
\end{equation}

Using~\eqref{eq:hExp} and~\eqref{eq:eExp} we can rewrite~\eqref{eq:IntSyndromeEntry} as
\begin{align}
 s_{i}^{(\w)}
 &=\sum_{j=1}^{n}h_{i,j}e_{j}^{(\w)}
 \\
 &=\sum_{j=1}^{n}\sum_{\ell=1}^{\Hdim}h_{i,j,\ell}\Hb_{\ell}\sum_{r=1}^{\Edim}e_{j,r}^{(\w)}\Eb_{r}
 \\
 &=\sum_{j=1}^{n}\sum_{\ell=1}^{\Hdim}\sum_{r=1}^{\Edim}h_{i,j,\ell}e_{j,r}^{(\w)}\Hb_{\ell}\Eb_{r}.\label{eq:expIntSyndromeProdBasis}
\end{align}

Defining 
\begin{equation}
	a_{i,\ell,r}^{(\w)}\defeq\sum_{j=1}^{n}h_{i,j,\ell}e_{j,r}^{(\w)},
\end{equation}
we can rewrite~\eqref{eq:expIntSyndromeProdBasis} as
\begin{equation}\label{eq:prodBasisExpInt}
	 s_{i}^{(\w)}=\sum_{\ell=1}^{\Hdim}\sum_{r=1}^{\Edim}a_{i,\ell,r}^{(\w)}\Hb_{\ell}\Eb_{r}.
\end{equation}
The coefficients $a_{i,\ell,r}^{(\w)}$ are from $\Fq$ and thus~\eqref{eq:prodBasisExpInt} shows that $s_{i}^{(\w)}$ can be written as an $\Fq$-linear combination of the elements
\begin{equation*}
  \{\Hb_{1}\Eb_{1},\Hb_{1}\Eb_{2},\dots,\Hb_{\Hdim}\Eb_{\Edim}\}
\end{equation*}
from the product space $\Pspace=\Hspace\Espace=\spannedBy{\{ab:a\in\Hspace,b\in\Espace\}}$. From~\eqref{eq:prodBasisExpInt} it follows that the syndrome space $\Sspace^{(\w)}=\spannedBy{s_{1}^{(\w)},s_{2}^{(\w)},\dots,s_{n-k}^{(\w)}}$ is a subspace of the product space $\Pspace$ which implies that $\SspaceInt \subseteq \Pspace$.

Since the error is taken randomly and the matrix $\H$ is full-rank, the syndrome entries $s_{i}^{(\w)}$ can be seen as random elements of $\Hspace\Espace$~\cite[Proposition 4.3]{gaborit2019LRPC}. Thus, the probability that $\SspaceInt \subsetneq \Pspace $ is equal to the probability that a random $(\Hdim\Edim)\times \u(n-k)$ matrix over $\Fq$ is not full-rank, which is $\leq q^{\Hdim\Edim - \u(n-k)}$, see~\cite[Lemma 4.4]{gaborit2019LRPC}.
\end{IEEEproof}

\subsubsection{Recovery of the Support of the Error}
Knowing the syndrome space $\SspaceInt$, we recover the support of the error in the second step of the algorithm.
\begin{lemma}
Let $\SspaceInt = \Pspace$ and $\SspaceInt_{\ell}=\spannedBy{\{\Hb_{\ell}^{-1}x:x\in \SspaceInt\}}$ for all $\ell\in\intervallincl{1}{\Hdim}$. Then
\begin{equation*}
\Espace\subseteq\left(\SspaceInt_{1}\cap\SspaceInt_{2}\cap\dots\cap\SspaceInt_{\Hdim}\right)
\end{equation*}
and
\begin{equation*}
\Pr[\Espace\subsetneq\left(\SspaceInt_{1}\cap\SspaceInt_{2}\cap\dots\cap\SspaceInt_{\Hdim}\right)] \leq \Edim q^{0.5\Hdim\Edim(\Hdim+1)-m}.
\end{equation*}
\label{lemma:recover_supp}
\end{lemma}

\begin{IEEEproof}
Since we assume that $\SspaceInt$ spans the whole product space $\Pspace$ we have that $\Eb_{r}\in\SspaceInt_{\ell}$ for all $r\in\intervallincl{1}{\Edim}$ which implies that $\Espace$ is a subspace of $\SspaceInt_{\ell}$ for all $\ell\in\intervallincl{1}{\Hdim}$.
Hence, we have 
\begin{equation*}
\Espace\subseteq\left(\SspaceInt_{1}\cap\SspaceInt_{2}\cap\dots\cap\SspaceInt_{\Hdim}\right)
\end{equation*}
and (see~\cite[Proposition 3.5]{gaborit2019LRPC})
\begin{equation*}
\Pr[\Espace\subsetneq\left(\SspaceInt_{1}\cap\SspaceInt_{2}\cap\dots\cap\SspaceInt_{\Hdim}\right)] \leq \Edim q^{0.5\Hdim\Edim(\Hdim+1)-m}.
\end{equation*}
\end{IEEEproof}
      
\subsubsection{Recovery of the Error Vector}
Once the support $\Espace$, i.e., $\Eb_{1},\hdots,\Eb_{\Edim}$, is known we determine the component errors $\vec{e}^{(\w)}$.
\begin{lemma}
Let $\Eb_1,\dots,\Eb_{\Edim}$ be a basis of the error and let $\Hdim \geq \frac{n}{n-k}$. Then an erasure decoder can determine the component errors $\vec{e}^{(\w)}$ with probability $\geq 1-\Edim q^{\Hdim\Edim-m}$.
\label{lemma:recover_vec}
\end{lemma}

\begin{IEEEproof}
By definition we have that
\begin{equation}
e_j^{(\w)} = \sum_{r=1}^{\Edim} e_{j,r}^{(\w)} \Eb_{r},\qquad \forall j\in[1,n],
\end{equation}
for $e_{j,r}^{(\w)} \in \Fq$ and $\w\in\intervallincl{1}{\u}$. We expand $s_i$ w.r.t. a product space basis and write~\eqref{eq:expIntSyndromeProdBasis} as
\begin{align}
s_i^{(\w)} &= \sum_{r=1}^{\Edim}\sum_{\ell=1}^{\Hdim} s_{i,\ell,r}^{(\w)} \Hb_{\ell}\Eb_{r} 
\\
&= \sum_{j=1}^{n}\sum_{\ell=1}^{\Hdim}\sum_{r=1}^{\Edim}h_{i,j,\ell} e_{j,r}^{(\w)}\Hb_{\ell}\Eb_{r}.
\end{align} 
To determine the error, we solve
\begin{equation}\label{eq:solveErrorInt}
s_{i,\ell,r}^{(\w)} =\sum_{j=1}^{n}h_{i,j,\ell} e_{j,r}^{(\w)},\quad
\begin{array}{ll}
 \ell\in[1,\Hdim], & r\in[1,\Hdim], \\
 i\in[1,n-k], & w\in[1,\u]
\end{array}
\end{equation}
for $e_{1,1}^{(1)}, e_{1,2}^{(1)}, \hdots, e_{n,\Edim}^{(\u)}$, which corresponds to~\cite[Equation~3]{gaborit2019LRPC}. 

Equation~\eqref{eq:solveErrorInt} corresponds to an inhomogeneous linear system of $(n-k)\u\Hdim\Edim$ equations in $\u n\Edim$ unknowns which can have a unique solution if $\u n\Edim \leq (n-k)\u\Hdim\Edim  \iff \Hdim \geq \frac{n}{n-k}$.

We get a unique solution if $\dim(\Espace\Hspace) = \Hdim\Edim$ which occurs with a probability $\geq 1-\Edim q^{\Hdim\Edim-m}$~\cite[Proposition 3.3]{gaborit2019LRPC}.
\end{IEEEproof}

The proposed decoding algorithm is summarized in Algorithm~\ref{alg:int_dec}.
We define $\Solve$ as a function that has as input the syndrome of an error, the parity-check matrix and the support of the error corresponding to the syndrome and returns the error corresponding to the syndrome. An efficient way of performing this step over $\Fq$ is shown in~\cite[Section 4.5]{gaborit2019LRPC}.

\begin{algorithm}[h!]
  \caption{Interleaved LRPC Decoding Algorithm}
  \label{alg:int_dec}
  \textbf{Input:}  $\H\in \Fqm^{(n-k)\times n}$, $\vec{y} = (\vec{y}^{(1)},\dots,\vec{y}^{(\u)})\in\Fqm^{\u n}$
  \\
    \textbf{Output:} $\vec{c} \in \Fqm^{\u n}$
    \begin{algorithmic}[1]
    \For{$\w \in [1,\u]$}
    \State $\vec{s}^{(\w)} \gets \vec{y}^{(\w)} \H^{\top} \in \Fqm^{(n-k)}$      
    \EndFor
    \State $\SspaceInt \gets \supp{\left((\s^{(1)}, \s^{(2)}, \dots, \s^{(\u)})\right)}$
    \For{$\ell \in [1,\Hdim]$}
    \State $\SspaceInt_{\ell} \gets \Hb_{\ell}^{-1} \SspaceInt$
    \EndFor
    \State $\Espace \gets \SspaceInt_{1} \cap \SspaceInt_{2} \cap \dots \cap \SspaceInt_{\Hdim} $
    \For{$\w \in [1,\u]$}
    \State $\e^{(\w)} \gets \Solve(\s^{(\w)},\H,\Espace) \in \Fqm^{n}$
    \EndFor
    \State \Return $\vec{c} = \vec{y} - \big(\e^{(1)},\dots,\e^{(\u)}\big)$
  \end{algorithmic}
\end{algorithm}

\begin{table}
\caption{Failure Probabilities of Interleaved Decoder (new)}
\renewcommand{\arraystretch}{1.3}
\begin{center}
\begin{tabular}{r|l}
Event & Probability \\
  \hline
  $\dim(\Hspace \Espace) < \Hdim\Edim$ & $\leq \Edim q^{\Hdim\Edim -m}$ \\
  $ \SspaceInt_{1} \cap \SspaceInt_{2} \cap \hdots \cap \SspaceInt_{\Hdim} \neq \Espace $& $\leq \Edim q^{0.5\Hdim\Edim(\Hdim+1)-m}$ \\
  $\dim(\SspaceInt)< \Hdim \Edim$ & $  \leq  q^{\Hdim\Edim - \u(n-k)}$ \\
\end{tabular}
\end{center}
\label{tab:failure}
\end{table}

\subsection{Upper Bounds on the Decoding Failure Rate}
As for the non-interleaved case, we have the condition $\Hdim \geq \frac{n}{n-k}$.
The component code needs to be constructed s.t. $\rk(\H_{\text{ext}}) = n $.

There are three events that make the proposed decoder fail:
\begin{enumerate}
\item $\dim(\Hspace \Espace) < \Hdim\Edim$
\item $\Espace \neq \SspaceInt_{1} \cap \SspaceInt_{2} \cap \hdots \cap \SspaceInt_{\Hdim}$
\item $\dim(\SspaceInt)< \Hdim \Edim$
\end{enumerate}

Upper bounds on the probabilities of the failure cases are derived in the Lemmas~\ref{lemma:comp_syndrome}, and Lemma~\ref{lemma:recover_vec}, respectively, and they are summarized in Table~\ref{tab:failure}.

We observe that the failure events 1) and 2) are not affected by interleaving. Condition 3) (which is usually the reason for a decoding failure~\cite[Section 4.3]{gaborit2019LRPC}) decreases exponentially in the interleaving order $\u$. 
The simulation results for different code parameters in Figure~\ref{fig:sim} show, that the derived upper bounds on the DFR provide a good estimate of the actual DFR.
A more detailed explanation of the simulation results is given in Section~\ref{sec:simResults}.

\subsection{Complexity Analysis}
The complexity of our proposed decoding algorithm of interleaved LRPC codes is derived in the following lemma.

\begin{lemma}
  The algorithm presented in Algorithm~\ref{alg:int_dec} requires
$O(\u n^2m^2)$
operations in $\Fq$ to decode a $[\u,\Hdim;n,k]$ interleaved LRPC code over $\Fqm$.
  \end{lemma}
\begin{IEEEproof}
Step 1) Computation of the syndrome space consists of computing $\u$ times a syndrome vector and determining a basis of the product space, which can be achieved by transforming a $m \times (n-k)\u$ matrix over $\Fq$ in reduced row echelon form. Computing the syndromes has a complexity of $O(\u(n-k)n)$ operations in $\Fqm$ or $O(\u(n-k)n m^2)\subset O(\u n^2 m^2)$ operations in $\Fq$~\cite[Remark 8]{puchinger2019HighOrderInt}, and the transformation in reduced row echelon form requires $O(\min\{m^2(n-k)\u;m(n-k)^2\u \})\subset O(\u \min\{m^2n;mn^2\})$. Thus for the first step, we require $O(\u n^2m^2)$ operations in $\Fq$.

Step 2) Recovery of the support of the error requires $O(4\Edim^2\Hdim^2m)$ operations in $\Fq$~\cite[Section 4.5]{gaborit2019LRPC}.

Step 3) Recovery of the error vector can be performed by solving $\u$ times a linear system of equations with $(n-k)\Hdim\Edim$ equations and $n\Edim $ unknowns over $\Fq$. This requires $O(\u(n-k)^2\Hdim^2\Edim^2n\Edim) \subset O(\u n^3\Edim^3\Hdim^2)$ operations over $\Fq$.
Alternatively, we can precompute a matrix $\D_{H} \in \Fq^{n\Edim\times n \Edim}$, as described in~\cite[Section 4.5]{gaborit2019LRPC}, and perform final step by $\u$ matrix--vector multiplications which requires $O(\u n^2 \Edim^2)$ operations over $\Fq$.
\end{IEEEproof}

We observe that the complexity of the first and the third step depends on the interleaving order $\u$ whereas the second step is independent of $\u$.

The results above results on decoding $\u$-interleaved LRPC codes are summarized in the following theorem.

\begin{theorem}[Decoding of Interleaved LRPC Codes]
 A $\u$-interleaved LRPC code $\mycode{IC}[\u,\Hdim;n,k]$ over $\Fqm$ can be decoded from an error of rank $t$ with probability a least
 \begin{equation}\label{eq:unionBound}
 	1-\left(\Edim q^{\Hdim\Edim -m}+\Edim q^{0.5\Hdim\Edim(\Hdim+1)-m}+q^{\Hdim\Edim - \u(n-k)}\right)
 \end{equation}
 requiring at most $O(\u n^2m^2)$ operations in $\Fq$.
\end{theorem}

\section{Evaluation of the Interleaved LRPC Decoder}

In this section we evaluate the proposed decoding algorithm for interleaved LRPC codes with respect to the error-correction capability, the computational complexity and the memory requirement for representing the code.  

For a fair comparison we consider a $[u,\Hdim;n,k]$ interleaved LRPC code $\Cint$ and compare it with a $[\Hdim;N,K]$ LRPC code $\Clong$ of length $N\defeq\u n$ and dimension $K:=\u k$ over the same field $\Fqm$.

\subsection{Error-Correction Capability}
Since $\frac{N}{N-K} = \frac{n}{n-k}$, we observe that the lower bound on $\Hdim$ is the same for $\Cint$ and $\Clong$. Further, since the upper bounds on the probability that $\dim(\Hspace \Espace) < \Hdim\Edim$ and $\Espace \neq \SspaceInt_{1} \cap \SspaceInt_{2} \cap \hdots \cap \SspaceInt_{\Hdim}$ are independent of $\u$, and $q^{\Hdim\Edim - (N-K)} = q^{\Hdim\Edim - \u(n-k)}$, the decoding failure probabilities for $\Cint$ and $\Clong$ are the same. 
This means that a $\u$-interleaved LRPC code $\Cint$ can correct the same number of errors as the LRPC code $\Clong$ with the same probability. This behavior can be observed in the simulation results given in Figure~\ref{fig:sim}.

\subsection{Computional Complexity}
A comparison of the computational complexity of the proposed decoding algorithm for $\Cint$ and the basic algorithm to decode $\Clong$ can be found in Table~\ref{tab:compl_comp}. 
We observe that the computational complexity of Step $1$ and Step $3$ of the decoding algorithm is reduced by a factor $\u$ and $\u^2$, respectively, by considering a $\u$-interleaved LRPC code instead of $\Clong$.
\begin{table}
\caption{Computational complexity in operations in $\Fq$.}
\renewcommand{\arraystretch}{1.3} 
\begin{center}
\begin{tabular}{l|l|l}
  & $\Cint$ & $\Clong$ \\
  \hline
1) Computation of $\Sspace$ & $O(\u n^2 m^2)$ &  $O(\u^2 n^2m^2)$\\
2) Recovery of $\Espace$ &  $O(4\Edim^2\Hdim^2m)$& $O(4\Edim^2\Hdim^2m)$ \\
3) Recovery of $\vec{e}$ & $O(\u n^2\Edim^2) $ & $O(\u^2 n^2\Edim^2) $ \\
\end{tabular}
\end{center}
\label{tab:compl_comp}
\end{table}

\subsection{Representation of the Generator Matrix} 
The generator matrix of the LRPC code $\Clong$ contains $\u^2 k n$ elements from $\Fqm$. 
The $\u$-interleaved LRPC code $\Cint$ has a generator matrix of the form
\begin{equation*}
  \Gint = \begin{pmatrix}
    \Gcomp &  \0 & \hdots & \0 \\
    \0&  \Gcomp & \hdots & \0 \\
    \vdots &\vdots & \ddots & \vdots \\
    \0&  \0 & \hdots & \Gcomp \\
    \end{pmatrix} \in \Fqm^{uk\times un}
\end{equation*} 
where $\Gcomp \in \Fqm^{k\times n}$ denotes the generator matrix of the component codes of $\Cint$.
Thus, we can interpret $\u$-interleaved LRPC codes as $[\Hdim;\u n,\u k]$ LRPC codes that have a special structure that permits an efficient representation of the generator matrix $\Gint$ requiring $kn$ elements from $\Fqm$. 
Hence, the amount of memory needed to represent the code is decreased by a factor of $\u^2$ compared to the non-interleaved case. 
The memory requirement can be further reduced by using $\u$-interleaved ideal LRPC codes.

\subsection{Simulation Results}\label{sec:simResults}

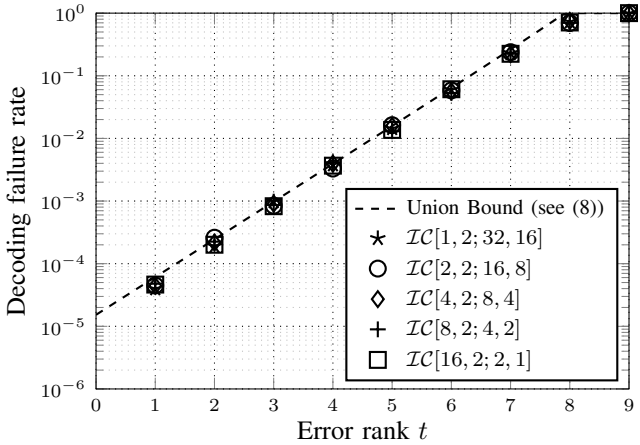
\begin{figure}
\begin{tikzpicture}
\begin{semilogyaxis}[
yticklabel style={/pgf/number format/fixed},
width=0.8\columnwidth,
height=5cm,
ymax = 1.0e-0,
ymin = 1.0e-6,
xmin = 0,
xmax = 9,
xtick={0,...,16},
ylabel={Decoding failure rate},
xlabel={Error rank $t$},
legend cell align=left,
grid=both,
minor grid style={dotted},
major grid style={dotted,black},
legend style={at={(0.98,0.02)},anchor=south east,thick,font={\footnotesize }},
]
\newcommand\pLambda{2}
\newcommand\pK{16}
\newcommand\pN{2*\pK}
\newcommand\pM{30}
\newcommand\pQ{2}




\addplot [domain=0:16, samples=100, black, dashed]{min(x* (\pQ^(\pLambda*x -\pM ))+x* (\pQ^(0.5*\pLambda*x*(\pLambda+1) -\pM ))+\pQ^(\pLambda*x-(\pN-\pK)),1)};
\addlegendentry{Union Bound (see~\eqref{eq:unionBound})};


\addplot[mark=star, black, only marks] table [x=t, y=fer, col sep=comma]{q2_m30_d2_n32_k16_ll1_id8662355210_results.txt};
\addlegendentry{$\mycode{IC}[1,2;32,16]$}

\addplot[mark=o, black, only marks] table [x=t, y=fer, col sep=comma] {q2_m30_d2_n16_k8_ll2_id1475666411_results.txt};
\addlegendentry{$\mycode{IC}[2,2;16,8]$}

\addplot[mark=diamond, black, only marks] table [x=t, y=fer, col sep=comma] {q2_m30_d2_n8_k4_ll4_id6399153220_results.txt};
\addlegendentry{$\mycode{IC}[4,2;8,4]$}

\addplot[mark=+, black, only marks] table [x=t, y=fer, col sep=comma] {q2_m30_d2_n4_k2_ll8_id7853823917_results.txt};  
\addlegendentry{$\mycode{IC}[8,2;4,2]$}

\addplot[mark=square, black, only marks] table [x=t, y=fer, col sep=comma] {q2_m30_d2_n2_k1_ll16_id6441436544_results.txt};
\addlegendentry{$\mycode{IC}[16,2;2,1]$}

\end{semilogyaxis}
\end{tikzpicture}	
\caption{Simulation results for different interleaving orders $\u$. The parameters are chosen such that all codes have the same rank $\lambda=2$, code rate $R=1/2$ and length $N=32$ over the field $\F_{2^{30}}$.}
\label{fig:sim}
\end{figure}

We performed simulations of $\u$-interleaved LRPC codes of different interleaving orders $\u \in \{1,2,4,8,16\}$.
All codes have the same rank $\Hdim = 2$, code rate $R=1/2$ and length $N=32$ over the field $\F_{2^{30}}$ ($q=2$ and $m=30$).
For each code, we generated one parity-check matrix for which we performed a Monte Carlo simulation and collect for each values of $t$, exactly $100$ decoding errors.
Note that the code with interleaving order $\u=1$ corresponds to a non-interleaved LRPC code.
The simulation results in Figure~\ref{fig:sim} show, that the union bound on the DFR in case of interleaving (see~\eqref{eq:unionBound}) give a good estimate of the measured DFR. 
Also we can observe that no loss due to interleaving occurs.

\section{Conclusion}

We proposed and analyzed an efficient decoding algorithm for horizontally $\u$-interleaved LRPC codes.
Upper bounds on the decoding failure probability as well as on the computational complexity were derived.
The results show that compared to a $\u$-times longer LRPC code of the same rank and code rate, the computational complexity is reduced by a factor of $\u$ for the same error-correction performance and decoding failure rate.
It was shown that interleaved LRPC codes admit a very compact representation of the code, which may be interesting for designing rank-metric code-based cryptosystems.
The proposed decoding algorithm may be further improved in terms of error-correction capability and decoding failure probability by using the ideas from the extended LRPC decoding algorithm in~\cite{gaborit2019LRPC}.


\section{Acknowledgments}

The authors would like to thank Antonia Wachter-Zeh for fruitful discussions and helpful comments. 

\bibliographystyle{IEEEtran}
\bibliography{main} 

\begin{thebibliography}{10}
\providecommand{\url}[1]{#1}
\csname url@samestyle\endcsname
\providecommand{\newblock}{\relax}
\providecommand{\bibinfo}[2]{#2}
\providecommand{\BIBentrySTDinterwordspacing}{\spaceskip=0pt\relax}
\providecommand{\BIBentryALTinterwordstretchfactor}{4}
\providecommand{\BIBentryALTinterwordspacing}{\spaceskip=\fontdimen2\font plus
\BIBentryALTinterwordstretchfactor\fontdimen3\font minus
  \fontdimen4\font\relax}
\providecommand{\BIBforeignlanguage}[2]{{%
\expandafter\ifx\csname l@#1\endcsname\relax
\typeout{** WARNING: IEEEtran.bst: No hyphenation pattern has been}%
\typeout{** loaded for the language `#1'. Using the pattern for}%
\typeout{** the default language instead.}%
\else
\language=\csname l@#1\endcsname
\fi
#2}}
\providecommand{\BIBdecl}{\relax}
\BIBdecl

\bibitem{Gabidulin_TheoryOfCodes_1985}
E.~M. Gabidulin, ``{Theory of Codes with Maximum Rank Distance},'' \emph{Probl.
  Inf. Transm.}, vol.~21, no.~1, pp. 3--16, 1985.

\bibitem{Delsarte_1978}
P.~Delsarte, ``{Bilinear Forms over a Finite Field with Applications to Coding
  Theory},'' \emph{J. Combin. Theory}, vol.~25, no.~3, pp. 226--241, 1978.

\bibitem{Roth_RankCodes_1991}
R.~M. Roth, ``{Maximum-Rank Array Codes and their Application to Crisscross
  Error Correction},'' \emph{IEEE Trans. Inf. Theory}, vol.~37, no.~2, pp.
  328--336, 1991.

\bibitem{silva2008rank}
D.~Silva, F.~R. Kschischang, and R.~Koetter, ``{A Rank-Metric Approach to Error
  Control in Random Network Coding},'' \emph{IEEE Trans. Inform. Theory},
  vol.~54, no.~9, pp. 3951--3967, 2008.

\bibitem{sidorenko2010decoding}
V.~Sidorenko and M.~Bossert, ``{Decoding Interleaved Gabidulin Codes and
  Multisequence Linearized Shift-Register Synthesis},'' in \emph{IEEE Int.
  Symp. on Inform. Theory (ISIT)}, 2010, pp. 1148--1152.

\bibitem{gabidulin2000space}
E.~M. Gabidulin, M.~Bossert, and P.~Lusina, ``{Space-Time Codes Based on Rank
  Codes},'' in \emph{IEEE Int. Symp. on Inform. Theory (ISIT)}, 2000, p. 284.

\bibitem{gaborit2013LRPC}
P.~Gaborit, G.~Murat, O.~Ruatta, and G.~Z{\'e}mor, ``{Low Rank Parity Check
  Codes and their Application to Cryptography},'' in \emph{Int. Workshop Coding
  Cryptogr. (WCC)}, vol. 2013, 2013.

\bibitem{Gabidulin2003Reducible}
E.~M. Gabidulin, A.~V. Ourivski, B.~Honary, and B.~Ammar, ``{Reducible Rank
  Codes and their Applications to Cryptography},'' \emph{IEEE Trans. Inf.
  Theory}, vol.~49, no.~12, pp. 3289--3293, 2003.

\bibitem{loidreau2016evolution}
P.~Loidreau, ``{An Evolution of GPT Cryptosystem}.''\hskip 1em plus 0.5em minus
  0.4em\relax ACCT, 2016.

\bibitem{faure2006new}
C.~Faure and P.~Loidreau, ``{A New Public-Key Cryptosystem Based on the Problem
  of Reconstructing p--Polynomials},'' in \emph{Coding and Cryptography}.\hskip
  1em plus 0.5em minus 0.4em\relax Springer, 2006, pp. 304--315.

\bibitem{wachter2018repairing}
A.~{Wachter-Zeh}, S.~{Puchinger}, and J.~{Renner}, ``{Repairing the
  Faure-Loidreau Public-Key Cryptosystem},'' in \emph{IEEE Int. Symp. on
  Inform. Theory (ISIT)}, June 2018, pp. 2426--2430.

\bibitem{gaborit2019LRPC}
\BIBentryALTinterwordspacing
N.~Aragon, P.~Gaborit, A.~Hauteville, O.~Ruatta, and G.~Z{\'{e}}mor, ``{Low
  Rank Parity Check Codes: New Decoding Algorithms and Applications to
  Cryptography},'' 2019. [Online]. Available:
  \url{http://arxiv.org/abs/1904.00357}
\BIBentrySTDinterwordspacing

\bibitem{rollo2019}
C.~{Aguilar Melchor}, N.~Aragon, M.~Bardet, S.~Bettaieb, L.~Bidoux, O.~Blazy,
  J.-C. Deneuville, P.~Gaborit, A.~Hauteville, A.~Otmani, O.~Ruatta, J.-P.
  Tillich, and G.~Z{\'e}mor, ``{ROLLO},'' \emph{https://www.pqc-rollo.org/}.

\bibitem{gaborit2017IBE}
P.~Gaborit, A.~Hauteville, D.~H. Phan, and J.-P. Tillich, ``{Identity-Based
  Encryption from Codes with Rank Metric},'' in \emph{Advances in Cryptology
  (CRYPTO 2017)}, J.~Katz and H.~Shacham, Eds., 2017, pp. 194--224.

\bibitem{Migler2004RankMatrix}
T.~{Migler}, K.~E. {Morrison}, and M.~{Ogle}, ``{{Weight and Rank of Matrices
  over Finite Fields}},'' \emph{arXiv Mathematics e-prints}, p. math/0403314,
  Mar 2004.

\bibitem{puchinger2019HighOrderInt}
\BIBentryALTinterwordspacing
S.~Puchinger, J.~Renner, and A.~Wachter{-}Zeh, ``{Decoding High-Order
  Interleaved Rank-Metric Codes},'' 2019. [Online]. Available:
  \url{http://arxiv.org/abs/1904.08774}
\BIBentrySTDinterwordspacing

\end{thebibliography}

\end{document}